Journal of Electrical Systems
and Information Technology

# Characterization of analog and digital control loops for bidirectional buck–boost converter using PID/PIDN algorithms

V. Viswanatha[1*], R. Venkata Siva Reddy[2] and Rajeswari[3]

*Correspondence:
viswas779@gmail.com
[1] Acharya Institute
of Technology, Bangalore,
India
Full list of author information
is available at the end of the
article

**Abstract**

This article presents the characterization of analog and digital control loops using PID/
PIDN control algorithms for bidirectional buck–boost converter (BBC). Control loops
of BBC are designed and implemented in MATLAB code using transfer functions in
time domain with unit step response and in frequency domain with bode plots and
pole-zero plots. These transfer functions are obtained by average large signal modeling
of BBC. Actions of analog and digital control loops are characterized in order to ensure
stability and dynamic response of BBC which is a bottleneck in renewable energy
applications. Improvement in dynamic response and stability of BBC with PIDN control
algorithm is demonstrated using bode plots, pole-zero plots, and step response. Con-
trol loop gain due to transfer functions of power stage and controllers is demonstrated,
and it is found stable in both analog and digital control loops. PIDN compensator is
proposed to maintain a healthy balance between the stability and transient behavior
since both are indirectly proportional. BBC is modeled using average large signal mod-
eling technique, simulated using MATLAB tool, and analysis of dynamic and stability
response is done through unit step input, bode plot, and pole-zero plot. Hardware is
designed and implemented using TMS320F28335 controller.

response, Analog and digital control loops, PID, PIDN

## Introduction

In today's technology worldwide, lots of changes are happening in almost all the
domains. Most of the domains run on electric power which is being generated using
sources like thermal, coal, nuclear, gas and so on. But these sources are non-renewable
sources that results in shortage of electric power in long run. Since the world is addicted
to usage of technology and if there is no power available over which technology runs,
world is going to end with traditional life style, and therefore the existing technology
is working on some findings using which electric power is generated from renewable
sources like solar energy, wind energy and fuel cell, etc. [1–3] Output of these sources
are DC power which can be stored using batteries when excess power is generated and
at the same time when no power is available then power stored in batteries can be taken
back for the loads also the energy generated by these renewable sources is fluctuating power and become unreliable system. This is the basic idea of finding of bidirectional DC–DC converter as power interface between main power and auxiliary energy storage system for efficient utilization power in the above said scenario [4–8]. The general architecture of bidirectional buck–boost converter in hybrid renewable energy system is shown in Fig. 1.

The BBC supposed to do the job of power conditioning among the loads, power storage systems, and DC bus. To do this, BBC has to be controlled in closed-loop system, and therefore, the closed-loop control mechanism needs to be tested with respect to stability and dynamic response.

Bode plots and pole-zero plots are the easiest tools to analyze the stability and dynamic response. Bode plots are used to measure the stability of such power converters in terms of gain margin, phase margin, damping factor, and bandwidth of control loops in frequency domain since time domain analysis is tedious [9]. There are two domains to develop and implement control loops for power converters: One is continuous time control loop as shown in Fig. 2a, and another one is discrete time control loop as shown in Fig. 2b.

When output voltage ($V_{OUT}$) is more than required magnitude, than the error voltage is going to be negative, compensator magnitude will increase that results in decreasing the duty cycle generated by the PWM function and the decreased duty cycle in turn results in reducing the output voltage. The same thing goes in opposite when output voltage is lower than the required magnitude. Control loops consist of voltage sensor and error signal generator; compensator and PWM function are having mainly three concerns: (1). Tracking which checks how close is the output to the commanded value, (2). Disturbance rejection which take care of how well does the output return to the proper value if the input voltage changes or the load changes or there is noise in the load signal voltage, and (3). Stability which ensures output signal does not run away.

In digital control loop, digital compensator is implemented. There are various control compensator algorithms like PID, PIDN, fuzzy PID, artificial neural network (ANN) and so on in terms of digital filters like IIR or FIR filters. Digital PWM is implemented with counter/timer/special hardware [10]. There is issue with digital control loop, i.e., delay time. Its time between sampling time and switching time.

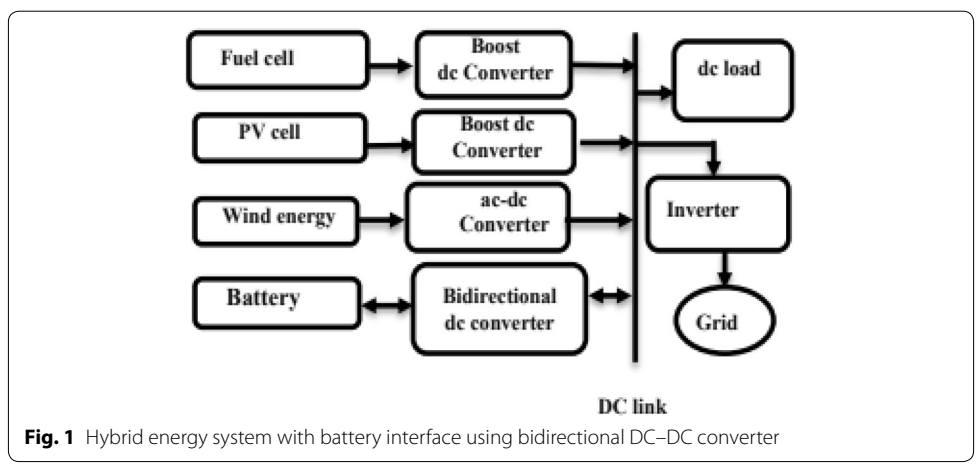

**Fig. 1** Hybrid energy system with battery interface using bidirectional DC–DC converter

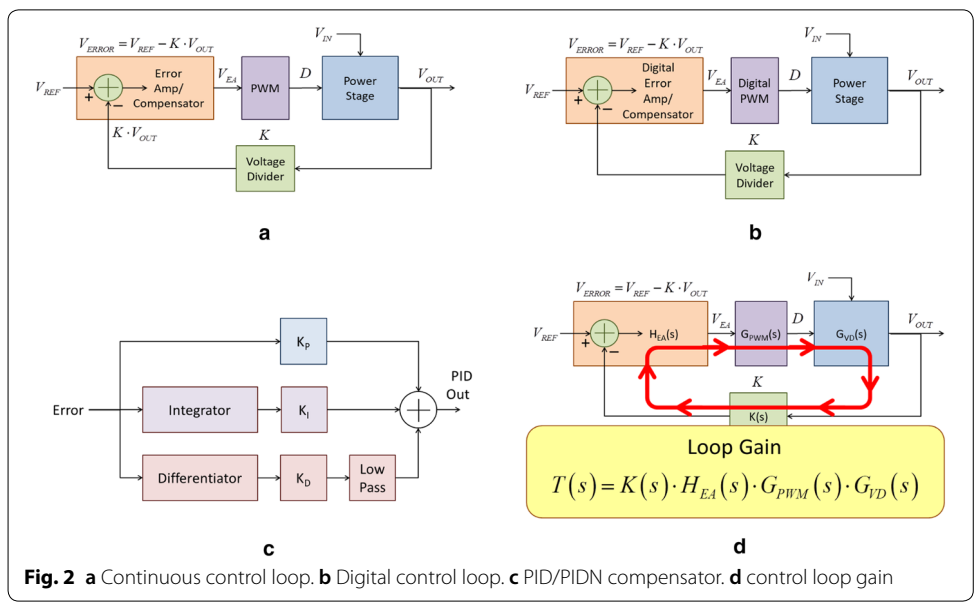

**Fig. 2** **a** Continuous control loop. **b** Digital control loop. **c** PID/PIDN compensator. **d** control loop gain

Delay time to be overcome such that sampling of output signal should happen at the middle of the off-time or middle of the on-time of PWM signal since switching instants are electrically very noisy [11].

Parallel PID/PIDN compensator is as shown in Fig. 2c. It takes the error signal ($V_{ERROR}$), and it is tuned by adjusting gain terms and generates the tuned signal ($V_{EA}$) which is input signal for PWM function. The tuned signal can either increase or decrease its magnitude based on positive or negative values of error signal in order to decide the stability and dynamic response of power stage. This compensator is good for one of the design where power stage characteristics are not well known.

The low-pass filter shown in Fig. 2c filter out the high-frequency-noise signal which is amplified by the differentiator. When the low-pass filter with huge value of filter coefficient ($N$) makes differentiator as pure differentiator, i.e., PID compensator that results in allowing high-frequency noise with control signal for power stage that in turn results in ripple in output signal [12]. When the value of filter coefficient ($N$) is nominal, i.e., equal to tuned value then the system is known as PID with filter called as PIDN.

The control loop of BBC either in analog or digital domain has gain known as loop gain $T(s)$ which can be determined as shown in Fig. 2d. Each block of control loop has its own transfer function. Gain of a signal which goes around the control loop is usually constant for feedback function $K(s)$ and PWM function $G_{pwm}(s)$. Power stage design sets $G_{VD}(s)$ and designer design compensator gain $H_{EA}(s)$ so that $T(s)$ gives desired dynamic response and stability [13].

For a control loop to be stable, loop gain is supposed to be $\leq 1$ at phase shift is $\leq 180°$. Also the phase at the crossover frequency indicates stability margin. Dynamic response is also determined with loop gain. Control loop has a faster transient response when the loop gain has higher crossover frequency where the loop gain crosses 0 dB. For fast response, preferred maximum phase shift is 135° at phase margin of 45°.

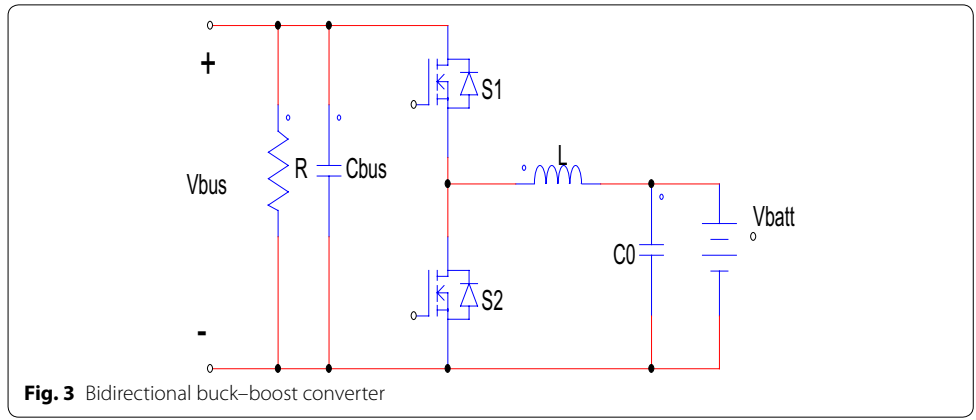

**Fig. 3** Bidirectional buck–boost converter

**Table 1  Specifications of BBC**

| Parameter | Value |
|---|---|
| DC bus voltage ($V_{bus}$) | 24 V |
| DC bus current ($I_{bus}$) | 3 A |
| Battery voltage ($V_{batt}$) | 12 V |
| Switching frequency ($F_s$) | 20 kHz |
| Load voltage ($V_0$) | 24 V |
| Load current ($I_0$) | 2.4 A |
| Duty cycle ($d$) | 0.5 |
| Inductor ($L$) | 1000 μH |
| DC bus capacitor ($C_{bus}$) | 250 μF |
| Load resistor ($R$) | 10 Ω |
| Capacitor across battery ($C_0$) | 125 μF |
| Battery resistance ($R_{batt}$) = $(N_s/N_p)*R_{inter}$ = (6/1)*30 m Ω | 0.18 Ω |

## Bidirectional DC–DC converter

The half bridge non-isolated bidirectional buck–boost converter as shown in Fig. 3 is taken into account in further design (as per the specification given in Table 1), modeling, and simulation and implementation process. It consists of two MOSFET switches with built-in anti-parallel diodes, one inductor, one 'R' load and two capacitors (one is at battery side and another one is at load side or DC bus side). It is a second-order system since two storage elements like inductor and capacitor come into picture when BBC operates in either buck mode or boost mode at given particular point of time. Inductor is common for both the modes; however, consideration of capacitor will change based on mode [14]. Based on the equivalent circuit due to operating mode, mathematical modeling is carried out using average large signal modeling technique. Switch 's2' is operating during boost mode and switch 's1' is non-operating. But in buck mode, it is vice versa. When switch 's2' is on, energy gets stored in the inductor by the virtue of current flowing through it. When 's2' is off, energy stored in inductor constitutes inductor voltage and supply voltage, i.e., battery voltage get added and will be fed to the DC bus through the in-built diode 'D1' of switch 's1' this is called boosting of voltage, and hence it is called boost mode of operation.

Buck mode is coming into picture when switch 's1' is operating and switch 's2' is non-operating. When switch 's1' is on, load, i.e., battery and source, i.e., DC bus are connected in series through energy storage element, i.e., inductor therefore the output voltage which is fed to the battery is less than the DC bus voltage [15].

## Mathematical modeling and control methods

To control the BBC in closed loop through modeling approach in analog and digital domain, PID and PIDN control methods are used and the method which gives best results and trade-off between stability and dynamic response are projected. Using average large signal modeling technique, The BBC shown in Fig. 3 is modeled, and it finds the application right from renewable energy systems to automobile systems. Later transfer functions in *S*-domain and *Z*-domain for each mode of operation are obtained based on the specification given in Table 1 by using MATLAB commands. Up next, based on transfer functions of power-stage of BBC, transfer functions in *S*-domain and *Z*-domains of control algorithms of PID and PIDN are obtained. At last, using transfer functions of power stage of BBC and control algorithms of PID and PIDN, closed-loop control responses are obtained in terms of step responses and bode plots in analog and digital domains using MATLAB commands. Simulation results are analyzed and compared w.r.t stability and dynamic responses in each mode of operation of BBC in both analog and digital domains in "Results and discussion" section.

During boost mode of operation of BBC, the state equation is shown by Eq. (1) and output equation is shown by Eq. (2).

$$\begin{bmatrix} x_1' \\ x_2' \end{bmatrix} = \begin{bmatrix} 0 & -\frac{(1-d)}{L} \\ \frac{(1-d)}{C_{\text{bus}}} & -\frac{1}{RC_{\text{bus}}} \end{bmatrix} \begin{bmatrix} x_1 \\ x_2 \end{bmatrix} + \begin{bmatrix} \frac{1}{L} \\ 0 \end{bmatrix} V_{\text{batt}}. \tag{1}$$

$$\begin{bmatrix} y_1 \\ y_2 \end{bmatrix} = \begin{bmatrix} 1 & 0 \\ 0 & 1 \end{bmatrix} \begin{bmatrix} i_L \\ V_C \end{bmatrix} + \begin{bmatrix} 0 \\ 0 \end{bmatrix} V_{\text{batt}}. \tag{2}$$

During buck mode of operation of BBC, the state equation is shown by Eq. (3) and output equation is shown by Eq. (4).

$$\begin{bmatrix} x_1' \\ x_2' \end{bmatrix} = \begin{bmatrix} 0 & -\frac{1}{L} \\ \frac{1}{C_0} & -\frac{1}{R_{\text{batt}}C_0} \end{bmatrix} \begin{bmatrix} x_1 \\ x_2 \end{bmatrix} + \begin{bmatrix} \frac{d}{L} \\ 0 \end{bmatrix} V_{\text{bus}}. \tag{3}$$

$$\begin{bmatrix} y_1 \\ y_2 \end{bmatrix} = \begin{bmatrix} 1 & 0 \\ 0 & 1 \end{bmatrix} \begin{bmatrix} i_L \\ V_C \end{bmatrix} + \begin{bmatrix} 0 \\ 0 \end{bmatrix} V_{\text{bus}}. \tag{4}$$

## Analog domain

Converter and controller models of BBC are designed in *S*-domain based on mode of operation, either buck mode or boost mode and type of control; either open loop or closed loop. Considering the design specification given in Table 1, the open-loop transfer function of boost mode (TF$_{\text{boost}}$) is obtained using Eqs. (1) and (2), and it is shown in Eq. (5) also the open-loop transfer function of buck mode (TF$_{buck}$) is obtained using (3) and (4), and it is shown in Eq. (6).

$$\text{TF}_{boost} = \frac{2 * 10^6}{S^2 + 400S + 1 * 10^6}.$$ (5)

$$\text{TF}_{buck} = \frac{4 * 10^6}{S^2 + 4.444 * 10^4 S + 8 * 10^6}.$$ (6)

The general equation for second-order open-loop transfer function is given by (6a).

$$G(s) = \frac{\omega_n^2}{s^2 + 2\xi\omega_n s}.$$ (6a)

Comparing Eq. (6a) with Eq. (5), the damping ratio is $\zeta = 0.1414$ and natural angular frequency $\omega_n = 1414.21$ rad/s. From this, it is clear that BBC in open-loop system under boost mode is underdamped system that results in oscillatory transient response of BBC at the natural frequency in open-loop control, i.e., without PID control, and it is as shown in Fig. 4.

Similarly, comparing Eq. (6a) with Eq. (6), the damping ratio is $\zeta = 11.11$ and natural angular frequency $\omega_n = 2000$ rad/s. From this, it is clear that BBC in open-loop system under buck mode is overdamped system, i.e., no oscillations, but it takes more time to reach stability during turn ON and turn OFF of the MOSFET switch (S1) used in buck mode. Since in this mode, battery is the load that offers negligible internal resistance, and hence, system behaves as if it is first-order system in open-loop control, i.e., without PID control, and it is as shown in Fig. 12.

The general equation for second-order closed-loop transfer function is given by (6b).

$$G(s) = \frac{\omega_n^2}{s^2 + 2\xi\omega_n s + \omega_n^2}.$$ (6b)

For boost mode, analog closed-loop controllers are designed with PID control who's transfer function is given in Eq. (7) and PIDN control who's transfer function is given in Eq. (8) with gain values which are obtained by autotuning of transfer function model of each mode of BBC in MATLAB tool.

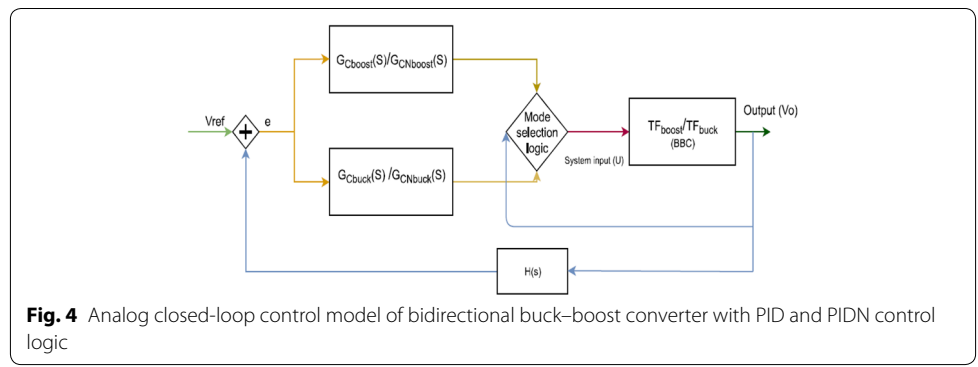

**Fig. 4** Analog closed-loop control model of bidirectional buck–boost converter with PID and PIDN control logic

PID control gains: kp = 4.63, ki = 1039.4, kd = 0.0048, $n$ = 1,703,100. PIDN control gains: kp = 4.63, ki = 1039.4, kd = 0.0048, $n$ = 17,031.

$$G_{Cboost}(S) = \frac{76947s^2 + 7.874 * 10^6 s + 1.77 * 10^9}{s^2 + 1.703 * 10^6 s}. \tag{7}$$

$$G_{CNboost}(S) = \frac{74.05s^2 + 7.977 * 10^4 s + 1.77 * 10^7}{s^2 + 1.703 * 10^4 s}. \tag{8}$$

For buck mode, analog closed-loop controllers are designed with PID control who's transfer function is given in Eq. (9) and PIDN control who's transfer function is given in Eq. (10) with gain values which are obtained by autotuning of transfer function model of each mode of BBC in MATLAB tool.

PID control gains: kp = 3.14, ki = 921.18, kd = − 0.0044, $n$ = 778,000. PIDN control gains: kp = 3.14, ki = 921.18, kd = − 0.0044, $n$ = 778.

$$G_{Cbuck}(S) = \frac{-3135s^2 + 2.443 * 10^6 s + 7.167 * 10^8}{s^2 + 7.78 * 10^5 s}. \tag{9}$$

$$G_{CNbuck}(S) = \frac{3.793 * 10^9 s^2 + 3364s + 7.169 * 10^5}{s^2 + 778.3s}. \tag{10}$$

$$M(s)_{CboostPID} = \frac{1.3894 * 10^{10}(s + 824.5)(s + 309)}{(s + 1.699 * 10^6)(s + 222.2)(s^2 + 4265s + 4.689 * 10^6)}. \tag{11}$$

$$M(s)_{CboostPIDN} = \frac{1.4809 * 10^8(s + 312.5)(s + 764.8)}{(s + 1.049 * 10^4)(s + 5280)(s + 1433)(s + 222.9)}. \tag{12}$$

$$M(s)_{CbuckPID} = \frac{-1.254 * 10^{10}(s - 1007)(s + 227.1)}{(s + 8.108 * 10^5)(s + 7648)(s + 3790)(s + 244)}. \tag{12a}$$

$$M(s)_{CbuckPIDN} = \frac{0.015171(s + 8.871 * 10^{11})(s + 213.1)}{(s + 4.427 * 10^4)(s + 221.3)(s^2 + 723.8s + 5.855 * 10^5)}. \tag{12b}$$

Equation (11) is a model represents closed-loop control of BBC in boost mode using PID controller in analog domain. This model is obtained by utilizing Eq. (5) represents converter being controlled in closed loop using Eq. (7) which represents controller.

Comparing equivalence of Eq. (11) with general equation of second-order system given in Eq. (6b) obtains $\zeta$ = 0.98 and natural angular frequency $\omega_n$ = 2165.4 rad/s. From this, it is clear that BBC in closed-loop system under boost mode using PID controller is underdamped system, and its step response is as shown in Fig. 8

This response has moderate oscillations compared to the response without PID control and it takes less time to reach stability as shown in Table 2 during turn ON and turn OFF of the MOSFET switch (S2) used in this mode.

**Table 2  Step response of BBC with PID, PIDN, and without PID in boost mode**

| BBC | Peak amplitude in volts | Overshoot (%) | Rise time in seconds | Settling time in seconds |
|---|---|---|---|---|
| Open loop | 3.05 | 52.7 | 0.00121 | 0.0196 |
| PID | 2.02 | 0.735 | 0.000416 | 0.0114 |
| PIDN | 2.08 | 4.1 | 0.000414 | 0.0115 |

Equation (12) is a model represents closed-loop control of BBC in boost mode using PIDN controller in analog domain. This model is obtained by utilizing Eq. (5) represents converter being controlled in closed loop using Eq. (8) which represents controller.

Comparing equivalence of Eq. (12) with general equation of second-order system given in Eq. (6b) obtains $\zeta = 1.46$ and natural angular frequency $\omega_n = 565.3$ rad/s. From this, it is clear that BBC in closed-loop system under boost mode using PIDN is overdamped system, and its step response is shown in Fig. 8.

Using PIDN controller, rise time and settling time is more compare to PID controller as shown in Table 2. Since the damping factor of PIDN-based closed-loop control of BBC is more than the damping factor of PID controller.

Buck mode models of closed-loop control of BBC are represented by Eqs. (12a) and (12b) using PID and PIDN controllers, respectively.

Converter being controlled in closed loop using Eq. (7) which represents controller.

Comparing equivalence of Eq. (12a) with general equation of second-order system given in Eq. (6b) obtains $\zeta = 2.09$ and natural angular frequency $\omega_n = 961.6$ rad/s. From this, it is clear that BBC in closed-loop system under buck mode using PID controller is overdamped system, and its step response is shown in Fig. 12.

This response has moderate oscillations compared to the response without PID control, and it takes less time to reach stability as shown in Table 5 during turn ON and turn OFF of the MOSFET switch(S1) used in this mode.

Equation (12b) is a model represents closed-loop control of BBC in buck mode using PIDN controller in analog domain. This model is obtained by utilizing Eq. (6) represents converter being controlled in closed loop using Eq. (10) which represents controller.

Comparing equivalence of Eq. (12b) with general equation of second-order system given in Eq. (6b) obtains $\zeta = 0.47$ and natural angular frequency $\omega_n = 765.179$ rad/s. From this, it is clear that BBC in closed-loop system under buck mode using PIDN is underdamped system, and its step response is shown in Fig. 12.

Analog closed-loop control architecture of BBC is shown in Fig. 4. Considering H(s) = unity. Open-loop and closed-loop models are implemented in MATLAB tool. Based on mode of operation, MATLAB code is developed for analog control of boost mode and buck mode of operation of BBC. Using Eqs. (5)–(10) a block diagram of the analog closed-loop control with PID and PIDN controller is obtained, and it is shown in Fig. 4.

### Digital domain

There are two basic methods to design a digital control for the system to be controlled. First one is redesign method and other one is direct method. In this work of develop of digital controller for BBC, redesign method is used. Digital control system with digital PID and PIDN controllers is developed separately for BBC [16].

Digital transfer functions of BBC based on modes of operation are shown in Eqs. (13) and (14). These models are obtained using Tustin transformation with sampling time '$t$' = 01 s, applied to boost mode and buck mode models of continuous systems given in Eqs. (5) and (6), respectively. Digital control system either open-loop or closed-loop control is stable when number of zeros should not be more than number of poles in the model otherwise at high frequencies, the gain of the system would be unbounded. The developed models represented in Eqs. (13)–(22) are not having zeros more than poles hence the developed models are stable digital systems. In addition, stability of these models are verified using bode plot and Z-plane in the coming section.

$$G_{\text{pboost}}(Z) = \frac{4.9899 * 10^{-5}(z+1)^2}{(z^2 - 1.996z + 0.996)}.$$
(13)

$$G_{\text{Pbuck}}(Z) = \frac{0.4501(z+1)^2}{(z+0.8007)(z+0.9991)}.$$
(14)

Transfer function of Digital PID and PIDN controllers for BBC in boost mode control is given in Eqs. (15) and (16), respectively.

$$G_{\text{Cboost}}(Z) = \frac{734.2z^2 - 1460z + 725.9}{z^2 - 0.2102z - 0.789}.$$
(15)

$$G_{\text{CNboost}}(Z) = \frac{68.6z^2 - 136.5z + 67.87}{z^2 - 1.843z + 0.8431}.$$
(16)

Transfer function of Digital PID and PIDN controllers for BBC in buck mode control is given in Eqs. (17) and (18), respectively.

$$G_{\text{Cbuck}}(Z) = \frac{49.12z^2 + 92.28z + 42.84}{z^2 - 5.141 * 10^{-5}z - 0.9999}.$$
(17)

$$G_{\text{CNbuck}}(Z) = \frac{49.12z^2 + 89.81z + 40.69}{z^2 - 0.05011z - 0.9499}.$$
(18)

$$M(z)_{\text{CboostPID}} = \frac{0.035341(z - 0.9918)(z - 0.9969)(z + 1)^2}{(z + 0.7889)(z - 0.9305)(z - 0.9887)(z - 0.9974)}.$$
(19)

Equation (19) is a model represents closed-loop control of BBC in boost mode using PID controller in digital domain. This model is obtained by utilizing Eq. (13) represents converter being controlled in closed loop using Eq. (15) which represents controller.

$$M(z)_{\text{CboostPIDN}} = \frac{0.0034115(z + 1)^2(z - 0.9969)(z - 0.9924)}{(z - 0.9899)(z - 0.9974)(z^2 - 1.839z + 0.851)}.$$
(20)

Equation (20) is a model represents closed-loop control of BBC in boost mode using PIDN controller in digital domain. This model is obtained by utilizing Eq. (13) represents converter being controlled in closed loop using Eq. (16) which represents controller.

$$M(z)_{\text{CbuckPID}} = \frac{0.95672(z + 0.8381)(z + 1)^2(z + 1.041)}{(z + 0.8791)(z + 0.9112)(z + 0.9985)(z + 1)}. \tag{21}$$

Equation (21) is a model represents closed-loop control of BBC in buck mode using PID controller in digital domain. This model is obtained by utilizing Eq. (14) represents converter being controlled in closed loop using Eq. (17) which represents controller.

$$M(z)_{\text{CbuckPIDN}} = \frac{0.95673(z + 0.8284)(z + 1)^3}{(z + 0.844)(z + 0.9991)(z^2 + 1.895z + 0.9009)}. \tag{22}$$

Equation (22) is a model represents closed-loop control of BBC in buck mode using PID controller in digital domain. This model is obtained by utilizing Eq. (14) represents converter being controlled in closed loop using Eq. (18) which represents controller.

Digital closed-loop control architecture of BBC is as shown in Fig. 5. Considering $H(Z) = 1$. Open-loop and closed-loop models are implemented in MATLAB tool. Based on mode of operation, MATLAB code is developed for digital control of boost mode and buck mode of operation of BBC. Using Eqs. (13)–(18) a block diagram of digital closed-loop control with PID and PIDN controller is obtained as shown in Fig. 5. From Fig. 5 it is understood that based on the value of load signal. The operating mode of BBC is changing between buck to boost mode and vice versa.

## Simulation

In order to simulate the BBC in closed-loop control with PID and PIDN control algorithms that are derived in "Mathematical modeling and control methods" section are implemented in MATLAB and solved using MATLAB facilities. The pseudocode for BBC in boost mode is shown in Fig. 6a and pseudocode for buck mode is shown in Fig. 6b. It consists of code for open-loop and closed-loop control of BBC using PID and PIDN algorithms in analog and digital domain. The program structure consists of converter parameters in terms of state equation (A & B) and output equation (C & D).From this, converter models in *S*-domain (Gs) and *Z*-domain (GsD) are obtained. Again from this, controller models of PID and PIDN are obtained in *S*-domain and *Z*-domain. By using converter model and controller model, closed-loop model is developed separately for PID control and PIDN control in *S*-domain and *Z*-domain. The MATLAB

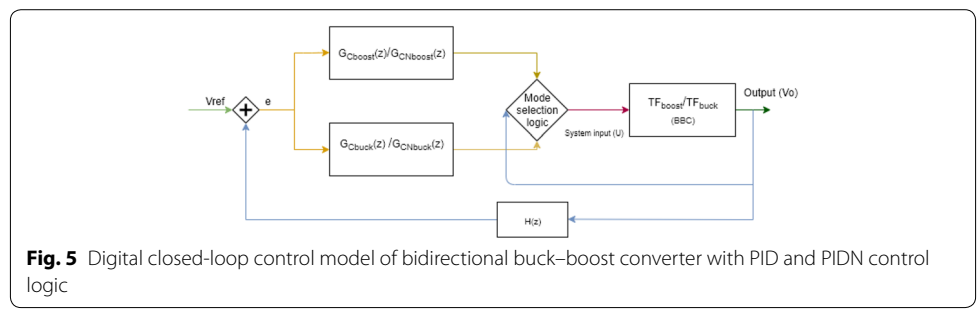

**Fig. 5** Digital closed-loop control model of bidirectional buck–boost converter with PID and PIDN control logic

**a**

```
********************************BBC in boost mode Code Begin*****************************
A=[0 -500 ; 2000 -400];
 B=[1000;0];
 C=[0 1];
 D=[0];
[b,a] =ss2tf(A,B,C,D)
mytf =tf(b,a) % transfer funtion of boost mode.
withoutpid=mytf
%step(withoutpid)
%pzmap(withoutpid)
 %bode(withoutpid)
hold on
Gs=zpk([],[(-200+979.79j),(-200-979.79j)],2000000)
kp=4.62284089651672
ki=1039.1837410487
kd=0.0040762390076835
contlr=tf([kd,kp,ki],[1,0])% transfer funtion of PID controller which is tuned in simulink for boost mode.
%PID=feedback(contlr*Gs,0.5)%value '0.5' is place since  duty cycle 1/2 is used in boost mode .
**%BBC-BOOST**************************************************************
GsD = c2d(Gs,0.1,'tustin')
digital_withoutpid=GsD
%bode(GsD)
%bode(digital_withoutpid)
% pzmap(digital_withoutpid)
 %PID with high value of 'n' become ideal PID, here value of 'n' considered is 100times more than the
 value obtained in tuning process .
 n=1703100.07
 analog_PID=tf([(kp+kd*n),(kp*n+ki),ki*n],[1,n,0])
 %analog_PID=tf([kd,kp,ki],[1,0])% transfer funtion of PID controller which is tuned in simulink for
 boost converter
 analog_Cloop=feedback(analog_PID*Gs,0.5)% gain 2,50% duty cycle
 %withpid=PID
 %pzmap( analog_PID)
 %bode( analog_PID)
hold on
%bode(analog_Cloop)
%pzmap( analog_Cloop)
%step(analog_Cloop)
DigitalPID = c2d(analog_PID,0.1,'tustin')
Digital_PID_Cloop=feedback(DigitalPID*GsD,1)
 %pzmap(DigitalPID)
 %bode(DigitalPID)
 %pzmap( DigitalPID)
 %step(analog_Cloop)
 %bode(Digital_PID_Cloop)
 %pzmap( Digital_PID_Cloop)
%PIDN*********************************************************************
 %PID with N(filter co-efficient)
kp=4.62284089651672
 ki=1039.1837410487
 kd=0.0040762390076835
 N=17031.07
 analog_PIDN=tf([(kp+kd*N),(kp*N+ki),(ki*N)],[1,N,0])% transfer funtion of PIDN controller which is tuned in simulink
 for boost converter
 %contlr=tf([392.1,8.912e05,2.366e08],[1,2.277e04,0])
 analog_CloopN=feedback(analog_PIDN*Gs,0.5)% analog closed loop control with PIDV....% gain 2,50% duty cycle
 %withpidn=PIDN
 %bode(analog_PIDN)
 %bode(analog_CloopN)
 %step(analog_CloopN)
 pzmap( analog_CloopN)
DigitalPIDN = c2d(analog_PIDN,0.1,'tustin')
Digital_PIDN_Cloop=feedback(DigitalPIDN*GsD,0.5)
 %bode(DigitalPIDN)
 %step(Digital_CloopN)
 %bode(Digital_PIDN_Cloop)
 %pzmap(Digital_PIDN_Cloop)
*****************************Code End***************************
```

**Fig. 6** **a** Mathematical model pseudocode for BBC in boost mode. **b** Mathematical model pseudocode for BBC in buck mode

instructions used for the development of such models are given in Fig. 6a which represents BBC control in boost mode control and Fig. 6b which represents BBC in buck mode control. The step response, bode plots, and pole-zero plots of these models are

**b**

```
************************BBC in Buck mode Code Begin*****************************
A=[0 -1000 ; 8000 -44444.45];
  B=[500;0];
  C=[0 1];
  D=[0];
  [b,a] =ss2tf(A,B,C,D)
  mytf =tf(b,a) % transfer funtion of buck converter
  analog_withoutpid=mytf
  %bode(analog_withoutpid)
  %pzmap(analog_withoutpid)
  Gs=zpk([],[(-180.7),(-44259.24)],4000000)
  kp=3.13926397775873
  ki=921.175630246025
  kd=-0.00403350179228654
  %contlr=tf([kd,kp,ki],[1,0])% transfer funtion of PID controller which is tuned in simulink for buck converter
  %PID=feedback(contlr*Gs,2)%value '2' is place since   duty cycle 1/2 is used in buck converer.
  %bode(withpid)
  %pzmap(withpid)
  GsD = c2d(Gs,0.1,'tustin')
  digital_withoutpid=GsD
  %bode(GsD)
  %bode(digital_withoutpid)
  pzmap(digital_withoutpid)
  %PID with high value of N become ideal PID, here value of N considered is 1000 times more than the value obtained in tuning process
  .%so  n=778000.297404
  n=778000.297404
  analog_PID=tf([(kp+kd*n),(kp*n+ki),ki*n],[1,n,0])
  %analog_PID=tf([kd,kp,ki],[1,0])% transfer funtion of PID controller which is tuned in simulink for buck converter
  analog_Cloop=feedback(analog_PID*Gs,2)% gain 2,50% duty cycle
  %withpid=PID
  %bode( analog_PID)
  %bodef( analog_PID)
  hold on
  %bode(analog_Cloop)
  %pzmap( analog_Cloop)
  %step(analog_Cloop)
  DigitalPID = c2d(analog_PID,0.1,'tustin')
  Digital_PID_Cloop=feedback(DigitalPID*GsD,1)
  %pzmap(DigitalPID)
  %bode(DigitalPID)
  %pzmap( analog_Cloop)
  %step(analog_Cloop)
  %bode(Digital_PID_Cloop)
  pzmap( Digital_Cloop)
  %PIDN*********************************************************
  %PID with N(filter co-efficient)
  kp=3.13926397775873
  ki=921.175630246025
  kd=-0.00403350179228654
  N=778.297404
  analog_PIDN=tf([(kp+kd*n),(kp*n+ki),(ki*N)],[1,N,0])% transfer funtion of PIDN controller which is tuned in simulink for buck
  converter
  %contlr=tf([392.1,8.912e05,2.366e08],[1,2.277e04,0])
  analog_CloopN=feedback(analog_PIDN*Gs,2)% anaog closed loop control with PIDV....% gain 2,50% duty cycle
  %withpidn=PIDN
  %bode(analog_PIDN)
  %bode(analog_CloopN)
  %step(analog_CloopN)
  %pzmap( analog_CloopN)
  DigitalPIDN = c2d(analog_PIDN,0.1,'tustin')
  Digital_PIDN_Cloop=feedback(DigitalPIDN*GsD,1)
  %bode(Digital_CloopN)
  %step(Digital_PIDN_Cloop)
  %bode(Digital_PIDN_Cloop)
  pzmap(Digital_CloopN)
  *******************************Code End*****************
```

**Fig. 6** continued

obtained in open-loop and closed-loop control and presented in "Results and discussion" section.

## Hardware

The hardware of bidirectional buck–boost converter which is controlled in closed loop using PIDN control algorithm with DSP controller (TMS320F28335) as shown in Fig. 7. It maintains the voltage stability using battery rated with 12 V, 7Ah in DC bus connected with renewable energy sources which provides unstable voltage for 57 Watt resistive load. Bidirectional buck–boost converter maintains stable voltage of 24 V and 2.4 A across DC bus with the help of battery by switching buck mode to boost mode and vice versa when any of the connected renewable energy sources gives variable voltage just because of variance in the sunlight/wind. The BBC works in buck mode when there is sufficient voltage on DC bus. In buck mode, BBC charges the battery with constant voltage of 12 V and 0.3 A. The BBC works in boost mode when there is no sufficient voltage on DC bus. In boost mode, BBC discharges the battery with constant voltage of 24 V and 2.4 A.

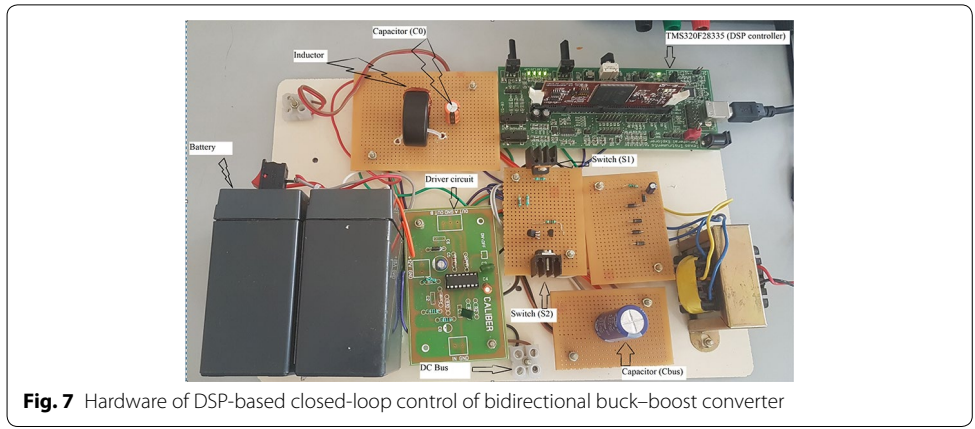

**Fig. 7** Hardware of DSP-based closed-loop control of bidirectional buck–boost converter

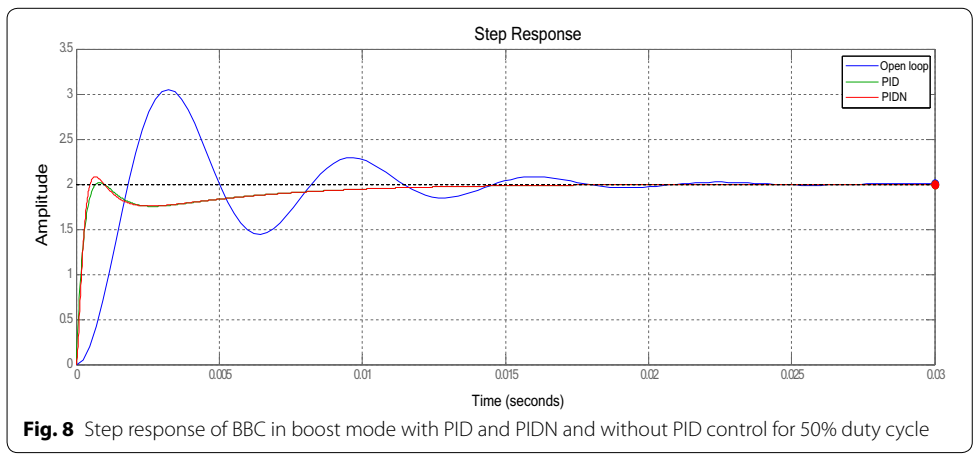

**Fig. 8** Step response of BBC in boost mode with PID and PIDN and without PID control for 50% duty cycle

## Results and discussion

The mathematical models of power stage and control stage of BBC in both open-loop and closed-loop control are developed, simulated, analyzed, and validated with respect to stability and dynamic response in MATLAB tool using PID and PIDN control laws in frequency domain as well as time domain. Since BBC works in two modes of operation for the purpose of power flow in both the directions, the determination of stability and dynamic response of each mode of operation is carried out separately.

(A). Stability and dynamic response in boost mode.

(B). Stability and dynamic response in buck mode.

A. Stability and dynamic response in boost mode

The validation of stability and dynamic response of BBC in boost mode operation is done through bode plots and pole-zero plots by making use of transfer function model given by Eq. (5) plots are developed using MATLAB code for the transfer functions. This feature becomes the base for the 1model to be embedded in design, simulation, and analysis. The approach of computational implementation through code is basically translates mathematical model to discrete programming code with various intermediate conversions in order to implement on hardware. The function ss2tf () does the job of converting the state space model of BBC in boost mode shown in Eqs. (1) and (2) into transfer function model in continuous frequency domain shown in Eq. (5) which is used to validate the step response in open loop and closed loop using PID and PIDN control in time domain. While bode plots and pole-zero plots in frequency domain measure the stability of both analog and digital control loops.

(i) Dynamic response

Transient response of BBC working in boost mode using unit step input is as shown in Fig. 8. From Fig. 8 various parameters are listed in Table 2. These parameters define the behavior of BBC for unit step input. If it is understood, then it is easy to understand the behavior of BBC for any type of input signal applied to it. PIDN controller has three control actions (proportional, integration and differentiation) along with filter operation which is associated with differentiator. Without filter with differentiator results in more noise termed as ripple in the load signal that will hinder the performance of the whole system. This is the issue with just PID control. The noise comes with sampled load signal gets amplified by differentiator therefore filter is connected in series with it. Better transient response can be obtained with just PID, whereas PIDN controller obtains better trade-off between stability and dynamic response. This is the major requirement of any system design.

(ii) Stability

It is easy to understand the variation of system parameters under frequency domain particularly the stability of closed-loop control of either analog or digital loops for BBC where the output voltage supposed to have a tight regulation for change in input voltage and load. This is possible with perfect control loop. Estimation of stability using various parameters in bode plots and pole-zero plots give the information that how the output voltage varies with the variation of input voltage, duty cycle, and the load with respect to frequency. Also estimation of damping factor from bode plots gives the information about reaction time of the system.

The frequency response of BBC in boost mode in analog domain is shown in Fig. 9 can be determined from transfer function using bode plots which is basically a graph of magnitude and phase of the transfer function as a function of frequency, where magnitude is plotted in decibels and phase in degrees. These plots reveal some key information about the control loop's performance. The first point of interest is the crossover frequency ($f_c$). Here, BBC in boost mode is showing 9.13 kHz under PIDN control. This is the frequency at which the control loop gain is unity (0 dB) and is also referred to as the loop bandwidth. The second point of interest is the place at which the phase lag reaches 180°. In this case, its infinity under PIDN control. The phase margin (PM) equals 180° minus the phase lag at $f_c$. In this case its 82.6°. The gain margin (GM) is the gain at a phase lag of 180°. In this case, its infinity under PIDN control. The system will be stable if the phase lag at $f_c$ is less than 180°. Here in buck mode under PIDN control, its 82.6° therefore it's stable. For most control loops, the engineers aim to achieve a PM greater than 45° and less than 180°.

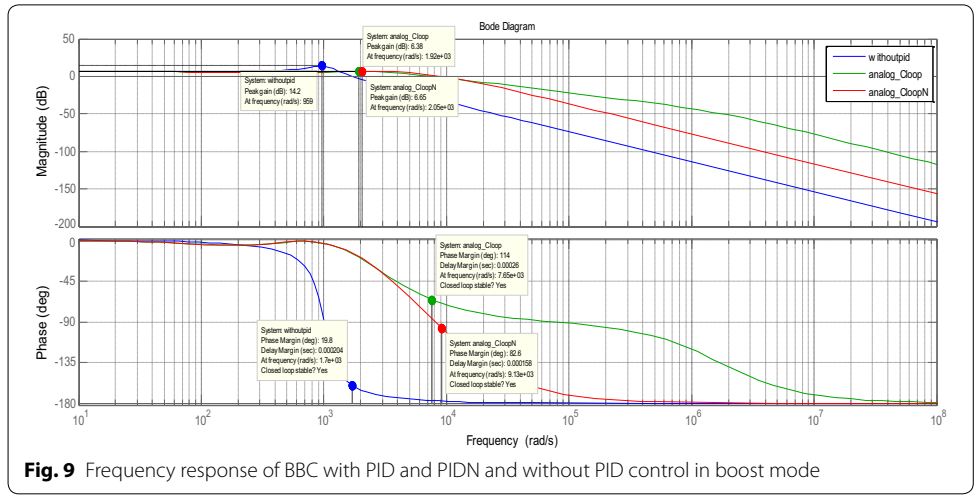

**Fig. 9** Frequency response of BBC with PID and PIDN and without PID control in boost mode

**Table 3 Comparison of stability analysis in analog domain of BBC in boost mode of operation**

| System | PM (degrees) | GM (db) | $\omega_g$ (rad/s) | $\omega_p$ (rad/s) | Delay margin (s) | Stability state |
|--------|-------------|---------|---------|---------|------------------|-----------------|
| Open loop (with-outpid) | 19.8 | Inf | Inf | 1700 | 0.000204 | Stable with poor dynamic response |
| PID control (analog_Cloop) | 114 | Inf | Inf | 7650 | 0.00026 | Stable with good dynamic response |
| PIDN control (analog_CloopN) | 82.6 | Inf | Inf | 9130 | 0.000158 | Better stable with good dynamic response |

Typically, a phase margin of 45° provides good transient response with good damping. For buck or boost switching system, the gain margin (GM) should be above 10 dB. In this case, GM is infinity. The data which defines the stability of BBC in boost mode with PID and PIDN control loops is extracted from bode plots of analog domain which is as shown in Fig. 9 and is tabulated in Table 3. Similarly, the data which defines the stability of BBC in boost mode with PID and PIDN control loops is extracted from bode plots of digital domain which is as shown in Fig. 10 and is tabulated in Table 4.

Pole-zero plot in analog domain for BBC in boost mode is demonstrated in open loop and closed loop using PID and PIDN as shown in Fig. 11a–c. From the figures, it is clear that BBC in boost mode is stable with various damping factors and overshoots in open-loop control as well as in closed-loop control since all the poles and zeros of all the systems are lying in the left-hand side of S-plane.

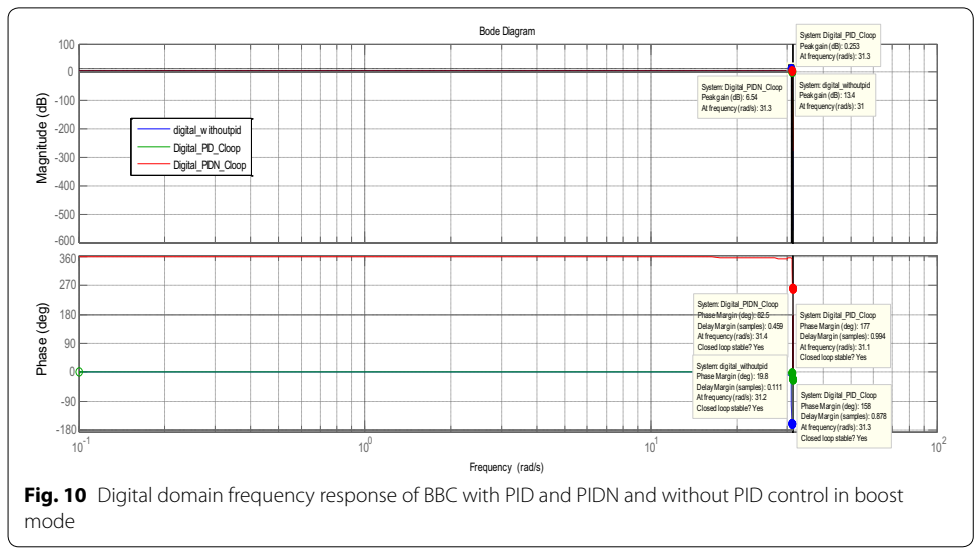

**Fig. 10** Digital domain frequency response of BBC with PID and PIDN and without PID control in boost mode

**Table 4** Comparison of stability analysis in digital domain of BBC in boost mode of operation

| System | PM (degrees) | GM (db) | $\omega_g$ (rad/s) | $\omega_p$ (rad/s) | Delay margin (samples) | Stability state |
|---|---|---|---|---|---|---|
| Open loop (digital_withoutpid) | 19.8 | Inf | Inf | 31.5 | 0.111 | Stable with poor dynamic response |
| PID control (digital_PID_Cloop) | 177 and 158 | Inf | Inf | 31.1 and 31.3 | 0.994 and 0.878 | Stable with good dynamic response |
| PIDN control (digital_PIDN_Cloop) | 82.5 | Inf | Inf | 31.4 | 0.459 | Better stable with good dynamic response |

Pole-zero plot in digital domain for BBC in boost mode is demonstrated in open loop and closed loop using PID and PIDN as shown in Fig. 11d. From the figure, it is clear that the BBC in boost mode is stable with various damping factors and overshoots in open-loop control as well as in closed-loop control since all the poles and zeros of all the systems are lying within the unit circle of Z-plane.

B. Stability and dynamic response in buck mode

Stability of the system is determined in frequency domain using transfer function of the same system. The transfer function for the BBC working in buck mode is developed using Eqs. (3) and (4) in MATLAB function ss2tf() and it is as shown in Eq. (6). Stability and dynamic response of BBC in buck mode of operation are characterized in frequency and time domain respectively using bode plot, pole-zero plot, and unit step response.

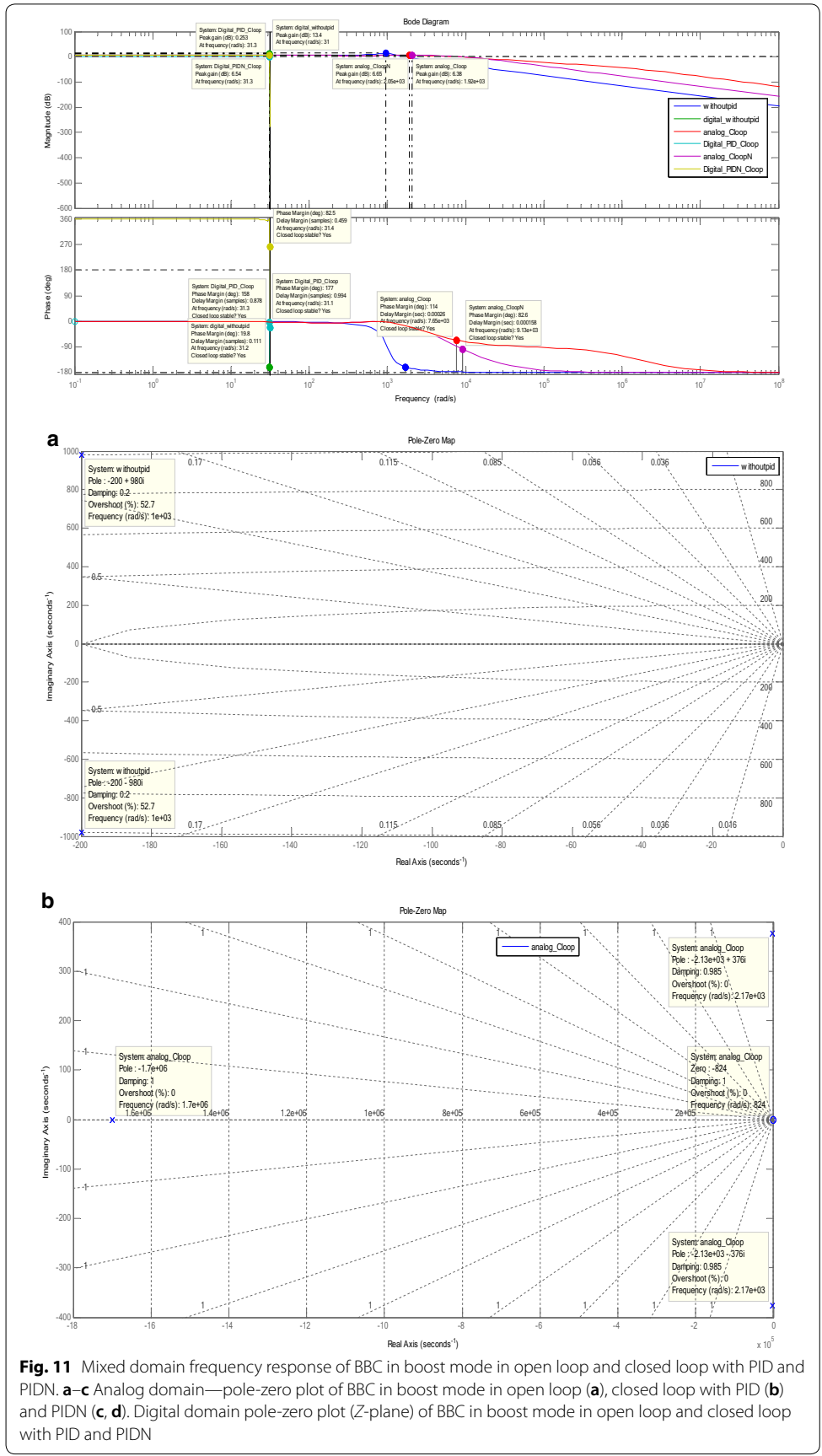

**Fig. 11** Mixed domain frequency response of BBC in boost mode in open loop and closed loop with PID and PIDN. **a–c** Analog domain—pole-zero plot of BBC in boost mode in open loop (**a**), closed loop with PID (**b**) and PIDN (**c, d**). Digital domain pole-zero plot (Z-plane) of BBC in boost mode in open loop and closed loop with PID and PIDN

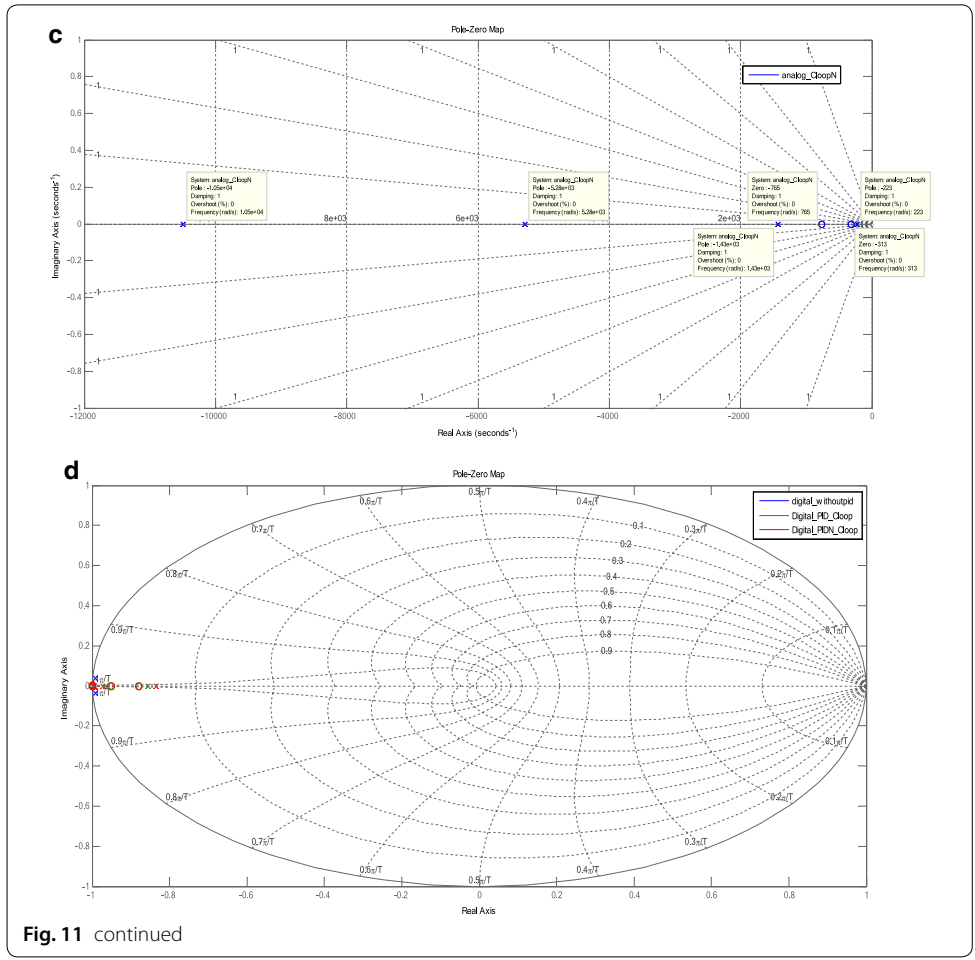

**Fig. 11** continued

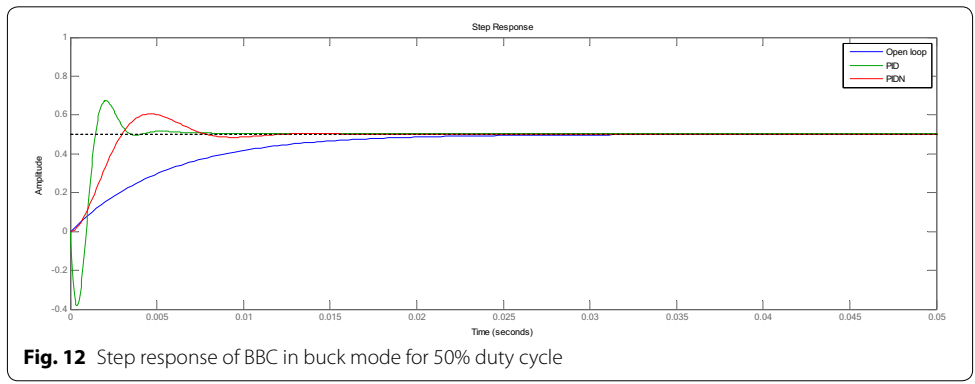

**Fig. 12** Step response of BBC in buck mode for 50% duty cycle

**Table 5 Step response of BBC with PID and PIDN and without PID in buck mode**

| BBC | Peak amplitude in volts | Overshoot | Rise time in seconds | Settling time in seconds |
|-----|--------------------------|-----------|----------------------|--------------------------|
| Open loop | 0.0% | 0.0% | 0.0122 | 0.0217 |
| PID | − 0.894 | 5.74 | 0.0000473 | 0.00237 |
| PIDN | 0.605 | 21.1 | 0.00201 | 0.0106 |

(i) Dynamic response

For BBC, dynamic response is coming into picture during switching of the switch 'S1' for buck mode of operation. To improve the dynamic response of BBC, PID and PIDN controllers are designed for BBC in both analog and digital domain. The dynamic response of BBC in buck mode is as shown in Fig. 12. Table 5 contains the characteristic properties of control loops for step input. Even in buck mode, PIDN gives better trade-off between stability and dynamic response.

(ii) Stability

It is easy to understand the variation of system parameters under frequency domain particularly the stability of closed-loop control of either analog or digital loops for BBC where the output voltage supposed to have a tight regulation for change in input voltage and load. This is possible with perfect control loop. Estimation of stability using various parameters in bode plots and pole-zero tools gives the information that how the output voltage varies with the variation of input voltage, duty cycle and the load with respect to frequency. Also estimation of damping factor from bode plots gives the information about reaction time of the system.

The frequency response of BBC in buck mode in analog domain is shown in Fig. 13 can be determined from transfer function using bode plots which is basically a graph of magnitude and phase of the transfer function as a function of frequency, where magnitude is plotted in decibels and phase in degrees. These plots reveal some key Information about the control loop's performance. The first point of interest is the crossover frequency ($f_c$). Here, BBC in buck mode is showing 360 Hz under PIDN control. This is the frequency at which the control loop gain is unity (0 dB) and is also referred to as the loop bandwidth. The second point of interest is the place at which the phase lag reaches 180°. In this case, its 5.74 kHz under PIDN control The phase margin (PM) equals 180° minus the phase lag at $f_c$. In this case, its 122°. The gain margin (GM) is the gain at a phase lag of 180°. In this case, its 40.8 dB under PIDN control. The system will be stable if the phase lag at $f_c$ is less than 180°. Here in buck mode under PIDN control, its 120°, and therefore, it is stable. For most control loops, the engineers aim to achieve a PM greater than 45° and less than 180°. Typically, a phase margin of 45° provides good transient response with good damping. For buck or boost switching system the gain margin should be above 10 dB. In this case, GM is 40.8 dB. The data which defines the stability of BBC in buck mode with PID and PIDN control loops is extracted from bode plots of analog domain which is shown in Fig. 13 and is tabulated in Table 6. Similarly, the data which defines the stability of BBC in buck mode with PID and PIDN control loops is extracted from bode plots of digital domain which is as shown in Fig. 14 and is tabulated in Table 7.

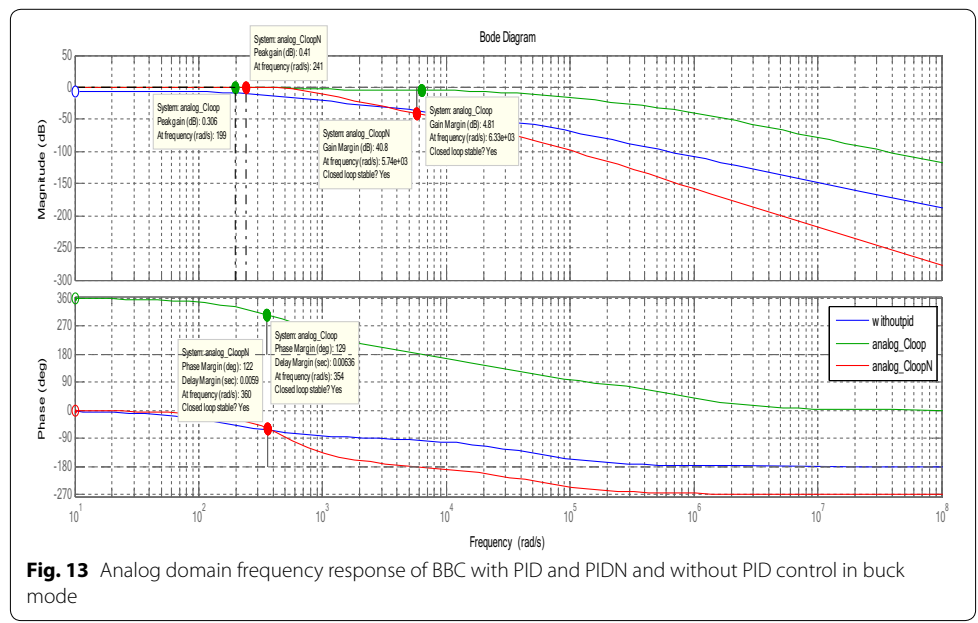

**Fig. 13** Analog domain frequency response of BBC with PID and PIDN and without PID control in buck mode

**Table 6** Comparison of stability analysis in analog domain of BBC in buck mode of operation

| System | PM (degrees) | GM (db) | $\omega_g$ (rad/s) | $\omega_p$ (rad/s) | Delay margin (s) | Stability state |
|---|---|---|---|---|---|---|
| Open loop (with-outpid) | Inf | Inf | Inf | Inf | Inf | Stable with poor dynamic response |
| PID control (analog_Cloop) | 129 | 4.81 | 6330 | 354 | 0.00636 | Stable with good dynamic response |
| PIDN control (analog_CloopN) | 122 | 40.8 | 5740 | 360 | 0.0059 | Better stable with good dynamic response |

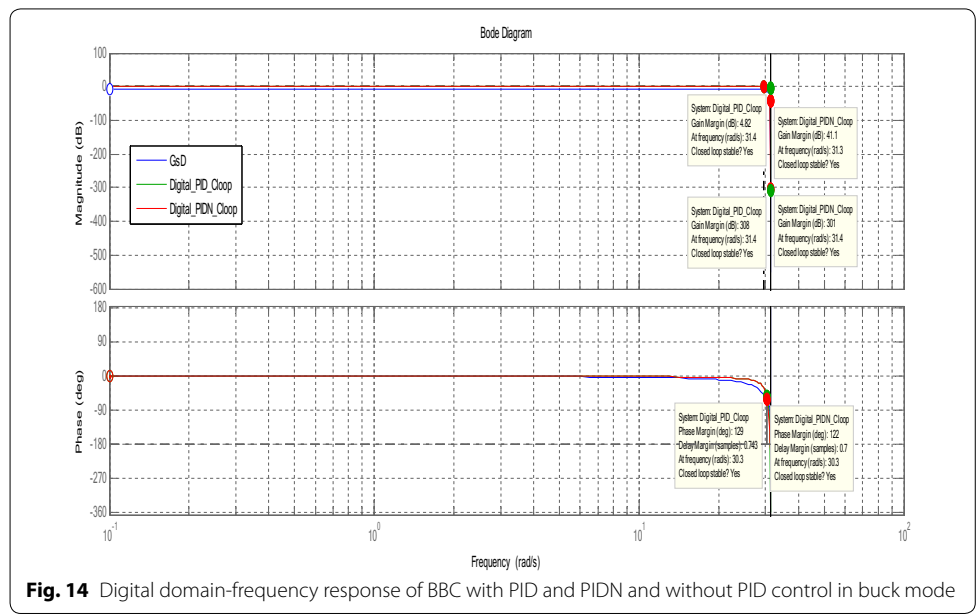

**Fig. 14** Digital domain-frequency response of BBC with PID and PIDN and without PID control in buck mode

From digital control technology, it is known that delay in control loop is more compare to analog control loop which offers minimum delay in loop, since the delay offered by ADC and DPWM is more in digital control that results in minimum control loop band [17, 18]. In hybrid renewable energy harvesting where power control system is used, the closed-loop control bandwidth plays an important role in the determination of sensitivity of power conditioning unit for transients which occur during switching instants [19–21]. But for the practical realization of advanced control algorithms cannot be possible with analog control systems. In addition to its environmental issues, it causes aging effects for analog control systems [22–25]. Using digital control systems, logic as well as input and output parameters can be easily modified as and when required [26]. Based on this knowledge, observing the data taken from analog and digital domain control of BBC through bode plots and listed in Tables 6 and 7.

**Table 7** Comparison of stability analysis in digital domain of BBC in buck mode of operation

| System | PM (degrees) | GM (db) | $\omega_g$ (rad/s) | $\omega_p$ (rad/s) | Delay margin (samples) | Stability state |
|---|---|---|---|---|---|---|
| Open loop (digital_withoutpid) | Inf | Inf | Inf | Inf | Inf | Stable with poor dynamic response |
| PID control (digital_PID_Cloop) | 129 | 4.82 and 308 | 31.4 | 30.3 | 0.743 | Stable with good dynamic response |
| PIDN control (digital_PIDN_Cloop) | 122 | 41.1 and 301 | 31.3 and 31.4 | 30.3 | 0.7 | Better stable with good dynamic response |

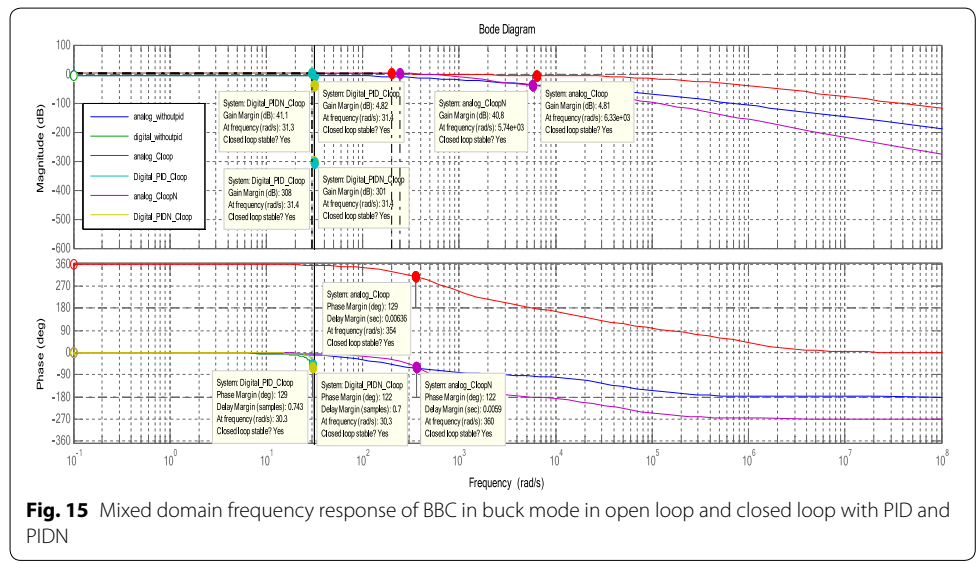

**Fig. 15** Mixed domain frequency response of BBC in buck mode in open loop and closed loop with PID and PIDN

Digital control loop offers more delay margin than analog control loop because of sampling process by ADC and quantization process by DPWM blocks in digital control loop [26]. These issues can be addressed using special type of control techniques like deadbeat digital control technique where sampling takes place for every switching cycle w.r.t. current considered as input to the loop from power stage [27]. There is a one more technique where multi-sampling of current is done by executing the control algorithm at frequency twice that of the converter frequency. This technique reduces DPWM delay for the maximum extent [28] (Fig. 15).

Pole-Zero plot in analog domain for BBC in buck mode is demonstrated in open loop and closed loop using PID and PIDN as shown in Figs. 16a–c. From the figures, it is clear that BBC in buck mode is stable with various damping factors and overshoots in open-loop control as well as in closed-loop control, all the poles and zeros of all the systems are lying in the left-hand side of S-plane.

Digital domain control gives more accuracy of control. Pole-zero plot in digital domain for BBC in buck mode is demonstrated in open loop and closed loop using PID and PIDN as shown in Fig. 16d. From the figure, it is clear that the BBC in buck mode is stable with various damping factors and overshoots in open-loop control as well as in closed-loop control with PIDN but not with PID control. One zero of PID is lying outside the unit circle since PID amplifies ripples along with load signal that will dilute the expected duty cycle of control signal in charging mode(buck mode). This issue is solved by PIDN control.

## Conclusion

Control loop gains of BBC in both analog and digital domain are characterized using bode plots and pole-zero plots and found to be less than unity, and it is much better with PIDN control law which filter out the noise due to switching exits in load signal.

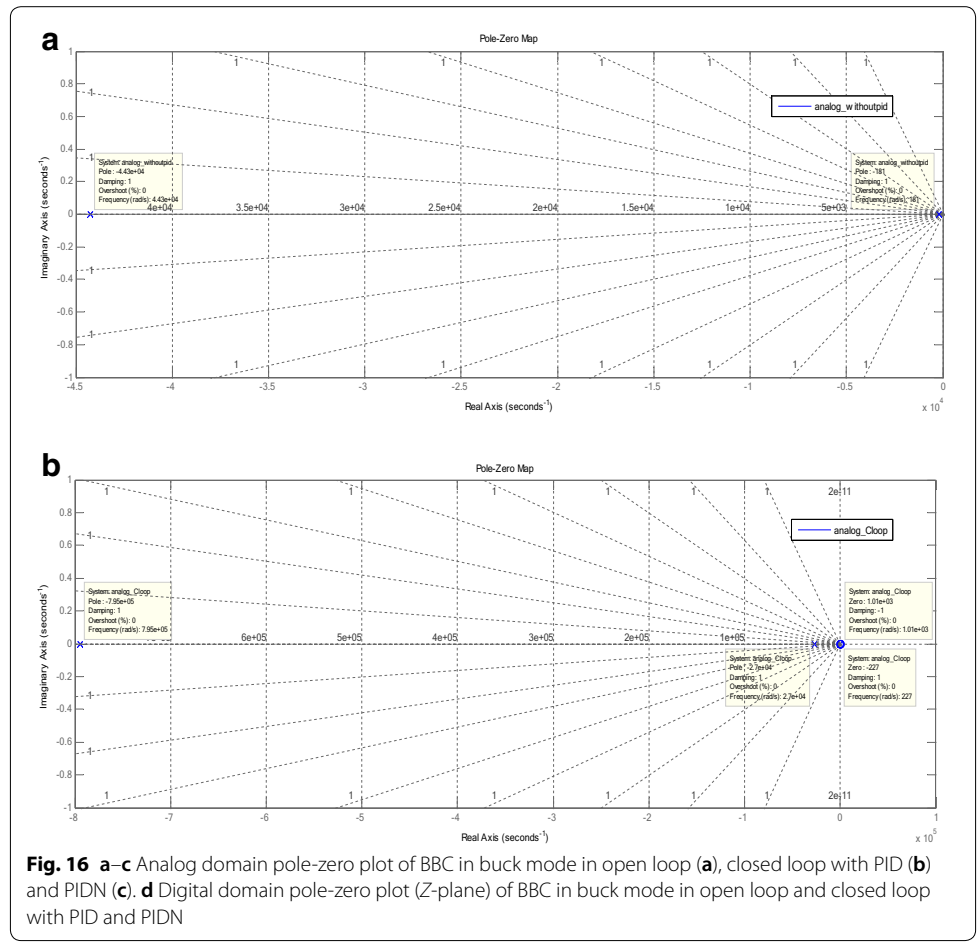

**Fig. 16** **a–c** Analog domain pole-zero plot of BBC in buck mode in open loop (**a**), closed loop with PID (**b**) and PIDN (**c**). **d** Digital domain pole-zero plot (*Z*-plane) of BBC in buck mode in open loop and closed loop with PID and PIDN

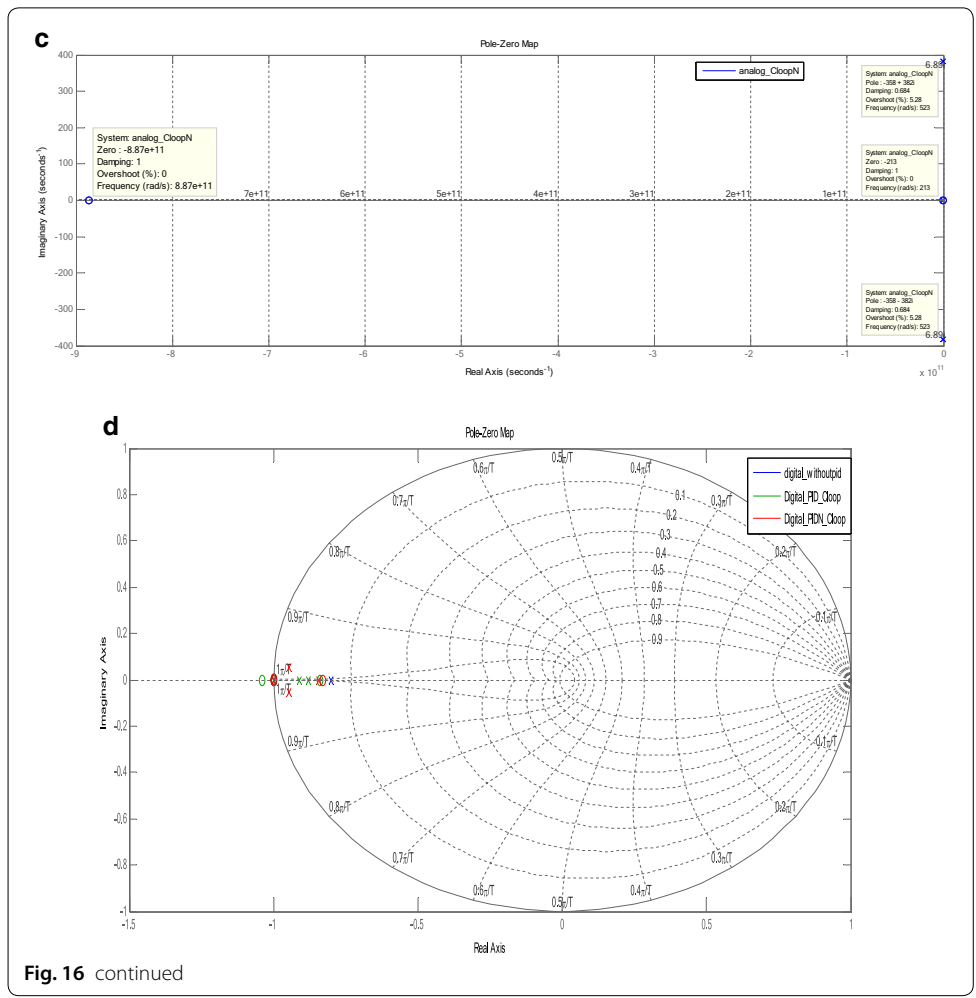

**Fig. 16** continued

Loop gain factor indicates that the system is stable with analog and digital control loops that results in good reliability of the system. Mathematical modeling is the key factor to develop stability and dynamic response analysis and correction can be incorporated for desired stability and dynamic response. Transfer functions of power stage are made to develop control loops which are implemented on digital signal processor (DSP) for the control of BBC prototype.

Further with same transfer function models of power stage, control algorithms like, adoptive PID, fuzzy logic with PIDN, model predictive control, intermediate control, neuro-FPGA and machine learning can be used to develop digital control loops which makes the control loop as smart control loop in digital control technology for power converters.

**Abbreviations**
BBC: Bidirectional buck–boost converter; PID: Proportional–integral derivatives; PIDN: Proportional–integral derivative with derivative filter order 'N'; ADC: Analog-to-digital converter; DPWM: Digital pulse-width modulation; DSP: Digital signal processing; PM: Phase margin; GM: Gain margin; HEVs: Hybrid electric vehicles; FCEVs: Fuel cell electric vehicles; $\omega_g$: Frequency response of gain margin; $\omega_p$: Frequency response of phase margin; inf: Infinity; ss2tf: State space to transfer function; $\zeta$: Damping ratio; $\omega_n$: Natural angular frequency; $TF_{boost}$: Transfer function of BBC in boost mode; $TF_{buck}$: Transfer function of BBC in buck mode; $G_{Cboost}(S)$: Transfer function of PID controller for BBC in boost mode; $G_{CNboost}(S)$:

Transfer function of PIDN controller for BBC in boost mode; $G_{Cbuck}(S)$: Transfer function of PID controller for BBC in buck mode; $G_{CNbuck}(S)$: Transfer function of PIDN controller for BBC in buck mode; $G_{Pboost}(Z)$: Digital transfer function of BBC in boost mode; $G_{Pbuck}(Z)$: Digital transfer function of BBC in buck mode; $G_{Cboost}(Z)$: Digital transfer function of PID controller for BBC in boost mode; $G_{CNboost}(Z)$: Digital transfer function of PIDN controller for BBC in boost mode; $G_{Cbuck}(Z)$: Digital transfer function of PID controller for BBC in buck mode; $G_{CNbuck}(Z)$: Digital transfer function of PIDN controller for BBC in buck mode; $M(s)_{CboostPID}$: Analog transfer function of closed-loop control of BBC using PID controller in boost mode; $M(s)_{CboostPIDN}$: Analog transfer function of closed-loop control of BBC using PIDN controller in boost mode; $M(s)_{CbuckPID}$: Analog transfer function of closed-loop control of BBC using PID controller in buck mod; $M(s)_{CbuckPIDN}$: Analog transfer function of closed-loop control of BBC using PIDN controller in buck mode; $M(z)_{CboostPID}$: Digital transfer function of closed-loop control of BBC using PID controller in boost mode; $M(z)_{CboostPIDN}$: Digital transfer function of closed-loop control of BBC using PIDN controller in boost mode; $M(z)_{CbuckPID}$: Digital transfer function of closed-loop control of BBC using PID controller in buck mode; $M(z)_{CbuckPIDN}$: Digital transfer function of closed-loop control of BBC using PIDN controller in buck mode.

## Acknowledgements

Not applicable.

## Authors' contributions

VV proposed the research point and contributed in the survey, data collection, deriving mathematical modeling, modeling of analog and digital control loops for BBC, simulation and derivation of results, technical review, submission of manuscript and applying the received review. VSRR contributed in the survey and data collection. He was also a major contributor in developing the MATLAB code for the mathematical models, implementation of hardware and analysis of results. R contributed in the design of hardware, analysis of results, writing the manuscript, grammatical and technical review. All authors read and approved the final manuscript.

## Author information

V. Viswanatha is presently working as Asst. Professor, Electronics and communication Engineering Department at Acharya Institute of Technology, Bangalore, India. He is pursuing his Ph.D. in the field of Embedded systems in power control and conversion. His areas of interests include IOT, Embedded & VLSI technology in power control and conversion.

R. Venkata Siva Reddy is presently working as Professor, Electronics and communication Engineering Department at REVA University Bangalore, India. He has completed his Ph.D. in the field of Digital System Design. His areas of interests include digital system design, signal processing and Systems & control engineering.

Rajeswari is presently working as Professor, Electronics and communication Engineering Department at Acharya Institute of Technology, Bangalore, India. She has completed her Ph.D. in the field of Signal Processing Engineering. Her areas of interests include signal processing, machine learning, deep learning and control engineering.

## Availability of data and materials

(1). Mathematical models of PID and PIDN control loops. (2). MATLAB code of control algorithms. (3) Simulation results and data obtained by hardware.

## Competing interests

The authors confirm that there are no known conflicts of interest associated with this publication and there has been no financial support for this work that could have influenced its outcome.

## Author details

[1] Acharya Institute of Technology, Bangalore, India. [2] Electronics and Communication Dept., REVA University, Bangalore, India. [3] Electronics and Communication Dept., Acharya Institute of Technology, Bangalore, India.

## References

1. Viswanatha V, Venkata Siva Reddy R (2018) Microcontroller based bidirectional buck–boost converter for photovoltaic power plant. J Electric Syst Inf Technol 5(3):745–758
2. Viswanatha V, Reddy RVS (2017) Digital control of buck converter using arduino microcontroller for low power applications. In: 2017 international conference on smart technologies for smart nation (SmartTechCon), Bangalore, 2017, pp 439–443
3. Viswanatha V, Venkata Siva Reddy R. Modeling, simulation and analysis of non-inverting buck-boost converter using PSIM. In: 2016 international conference on circuits, controls, communications and computing (I4C), Bangalore, 2016, pp 1–5
4. Ravi D et al (2018) Bidirectional DC to DC converters: an overview of various topologies, switching schemes and control techniques. Int J Eng Technol. 7(4):360–365
5. Caricchi F, Crescimbini F, Capponi FG, Solero L (1998) Study of bi-directional buck-boost converter topologies for application in electrical vehicle motor drives. APEC'98 Thirteen Annu Appl Power Electron Conf Expo 1:287–293
6. Lai J-S, Nelson DJ (2007) Energy management power converters in hybrid electric and fuel cell vehicles. Proc IEEE 95(4):766–777
7. Emadi A, Williamson SS, Khaligh A (2006) Power electronics intensive solutions for advanced electric, hybrid electric, and fuel cell vehicular power systems. IEEE Trans Power Electron 21(3):567–577

8.  Plesko H, Biela J, Luomi J, Kolar JW (2008) Novel concepts for integrating the electric drive and auxiliary DC–DC converter for hybrid vehicles. IEEE Trans Power Electron. 23(6):3025–3034

9.  Gorji SA, Ektesabi M, Zheng J. Double-input boost/Y-source DC–DC converter for renewable energy sources. In: Proc. IEEE SPEC '16, 2016, Auckland, pp 1–6

10. Thummala P, Maksimovic D, Zhang Z, Andersen MAE (2016) Digital Control of a High-Voltage (2.5 kV) Bidi-rectional DC–DC Flyback Converter for Driving a Capacitive Incremental Actuator. IEEE Trans Power Electron 31(12):8500–8516

11. Amjadi Z, Williamson SS (2014) Digital control of a bidirectional DC/DC switched capacitor converter for hybrid electric vehicle energy storage system applications. IEEE Trans Smart Grid 5(1):158–166

12. Baek J, Choi W, Cho B (2013) Digital adaptive frequency modulation for bidirectional DC–DC converter. IEEE Trans Ind Electron 60(11):5167–5176

13. Guida B, Rubino L, Marino P, Cavallo A (2010) Implementation of control and protection logics for a bidirectional DC/DC converter. In: Proc IEEE Int Symp Ind Electron. 2010, Bari, pp 2696–2701

14. Cornea O, Andreescu G, Muntean N, Hulea D (2017) Bidirectional power flow control in a DC microgrid through a switched-capacitor cell hybrid DC–DC converter. IEEE Trans Ind Electron 64(4):3012–3022

15. Forouzesh M, Siwakoti YP, Gorji SA, Blaabjerg F, Lehman B (2017) Step-up DC–DC converters: a comprehensive review of voltage-boosting techniques, topologies, and applications. IEEE Trans Power Electron 32(12):9143–9178

16. Jung J, Kim H, Ryu M, Baek J (2013) Design methodology of bidirectional CLLC resonant converter for high-fre-quency isolation of DC distribution systems. IEEE Trans Power Electron 28(4):1741–1755

17. Corradini L, Mattavelli P (2008) Modeling of multisampled pulse width modulators for digitally controlled DCDC converters. IEEE Trans Power Electron 23(4):1839–1847

18. Ilic M, Maksimovic D (2008) Digital average current-mode controller for DC–DC converters in physical vapor deposi-tion applications. IEEE Trans Power Electron 23(3):1428–1436

19. Ramrez-Murillo H, Restrepo C, Konjedic T, Calvente J, Romero A, Baier CR, Giral R (2018) An efficiency comparison of fuel-cell hybrid systems based on the versatile buck–boost converter. IEEE Trans Power Electron 33(2):1237–1246

20. Ramrez-Murillo H, Restrepo C, Calvente J, Romero A, Giral R (2015) Energy management of a fuel-cell serial–parallel hybrid system. IEEE Trans Ind Electron 62(8):5227–5235

21. Ramírez-Murillo H et al (2014) Energy management dc system based on current-controlled buck-boost modules. IEEE Trans Smart Grid 5(5):2644–2653

22. Jones DC, Erickson RW (2013) A nonlinear state machine for dead zone avoidance and mitigation in a synchronous noninverting buck boost converter. IEEE Trans Power Electron 28(1):467–480

23. Restrepo C, Konjedic T, Member S, Calvente J, Giral R, Member S (2015) Hysteretic transition method for avoid-ing the dead-zone effect and subharmonics in a non-inverting buck-boost converter. IEEE Trans Power Electron 30(6):3418–3430

24. Zhang N, Zhang G, See KW (2018) Systematic derivation of deadzone elimination strategies for the noninverting synchronous buckboost converter. IEEE Trans Power Electron 33(4):3497–3508

25. Callegaro L, Ciobotaru M, Pagano DJ, Turano E, Fletcher JE (2018) A simple smooth transition technique for the noninverting buckboost converter. IEEE Trans Power Electron 33(6):4906–4915

26. Vidal-Idiarte E, Marcos-Pastor A, Giral R, Calvente J, Martinez-Salamero L (2017) Direct digital design of a sliding mode-based control of a PWM synchronous buck converter. IET Power Electron 18(1):411–419

27. Chen J, Prodic A, Erickson RW, Maksimovic D (2003) Predictive digital current programmed control. IEEE Trans Power Electron 18(1):411–419

28. Restrepo C, Konjedic T, Flores-Bahamonde F, Vidal-Idiarte E, Calvente J, Giral R (2019) Multisampled digital average current controls of the versatile buck-boost converter. IEEE J Emerg Sel Topics Power Electron 7(2):879–890

## Publisher's Note

Springer Nature remains neutral with regard to jurisdictional claims in published maps and institutional affiliations.